\documentclass[aps,physrev,12pt,preprint,groupedaddress]{revtex4-2}

\usepackage{lmodern}
\usepackage{dcolumn}
\usepackage{amsmath, amsthm, amssymb, amsfonts}

\newcommand*{\defeq}{\mathrel{\vcenter{\baselineskip0.5ex \lineskiplimit0pt
                     \hbox{\scriptsize.}\hbox{\scriptsize.}}}%
                     =}

\swapnumbers 

\newtheoremstyle{def}%
    {3pt}{3pt}%
    {}{}%
    {\bfseries}{.}%
    { }%
    {\thmname{#1}\thmnumber{ #2}\thmnote{ (#3)}}
\theoremstyle{def}

\newtheoremstyle{thm}%
    {3pt}{3pt}%
    {\itshape}{}%
    {\bfseries}{.}%
    { }%
    {\thmname{#1}\thmnumber{ #2}\thmnote{ (#3)}}
\theoremstyle{thm}

\newtheoremstyle{remspace}
    {3pt}{3pt}%
    {}{}%
    {\itshape}{.}%
    { }%
    {\thmname{#1}\thmnumber{ #2}\thmnote{ (#3)}}
\theoremstyle{remspace}

\swapnumbers
\theoremstyle{remark}

\everymath=\expandafter{\the\everymath\displaystyle}
\IfFileExists{scrextend.sty}{
\usepackage[fontsize=10.000000pt]{scrextend}
}{
\renewcommand{\normalsize}{\fontsize{10.000000}{12.000000}\selectfont}
\normalsize
}
   
\makeatletter\@ifpackageloaded{underscore}{}{\usepackage[strings]{underscore}}\makeatother

\usepackage{comment}
\usepackage{hyperref}
\usepackage{cleveref}
\usepackage{braket, bm}
\usepackage{thmtools}
\usepackage{graphicx}\graphicspath{ {plots/} }
\usepackage{tikz}
\usepackage{subcaption}
\usepackage[english]{babel}
\usepackage[dvipsnames]{xcolor}

\begin{document}

\title{\Large
Benchmarking Lie-Algebraic Pretraining and Non-Variational QWOA for the MaxCut Problem
}

\author{\large Matthaus Zering}
\email[]{matthaus.zering@research.uwa.edu.au}

\author{Jolyon Joyce}
\email[]{jolyon.joyce@uwa.edu.au}
\author{Tal Gurfinkel}
\email[]{tal.gurfinkel@uwa.edu.au}
\author{Jingbo Wang}
\email[]{jingbo.wang@uwa.edu.au}

\affiliation{\large Department of Physics, The University of Western Australia, Perth, WA 6009, Australia}

\date{\today}

\begin{abstract}
The Quantum Approximate Optimization Algorithm (QAOA) is a leading candidate for achieving quantum advantage in combinatorial optimization on Near-Term Intermediate-Scale Quantum (NISQ) devices. However, random initialization of the variational parameters typically leads to vanishing gradients, rendering standard variational optimization ineffective. This paper provides a comparative performance analysis of two distinct strategies designed to improve trainability: Lie algebraic pretraining framework that uses Lie-algebraic classical simulation to find near-optimal initializations, and non-variational QWOA (NV-QWOA) that targets a restrict parameter subspace covered by 3 hyperparameters. We benchmark both methods on the unweighted Maxcut problem using a circuit depth of $p = 256$ across 200 Erdős-Rényi and 200 3-regular graphs, each with 16 vertices. Both approaches significantly improve upon the standard randomly initialized QWOA. NV-QWOA attains a mean approximation ratio of 98.9\% in just 60 iterations, while the Lie-algebraic pretrained QWOA improves to 77.71\% after 500 iterations. That optimization proceeds more quickly for NV-QWOA is not surprising given its significantly smaller parameter space, however, that an algorithm with so few tunable parameters reliably finds near-optimal solutions is remarkable. These findings suggest that the structured parameterization of NV-QWOA offers a more robust training approach than pretraining on lower-dimensional auxiliary problems. Future work is needed to confirm scaling to larger problem sizes and to asses generalization to other problem classes.

\end{abstract}
\maketitle

\section{Introduction}\label{sec:intro}

Combinatorial optimization problems are foundational to a wide range of real-world problems and are generally hard to solve. They are also one of the leading areas in which quantum computers may offer a potential advantage, particularly on near term hardware~\cite{preskill2018nisq}. A pioneering algorithm in this area is the Quantum Approximate Optimization Algorithm (QAOA), which arrives at approximate solutions through varying the parameters of a parameterized quantum circuit (PQC). Variational quantum algorithms (VQAs), such as QAOA, are considered to be a leading candidate for implementation on noisy intermediate-scale quantum (NISQ) hardware and beyond (\cite{preskill2018nisq,Biamonte2017QuantumML,blekos2024reviewQAOA}). QAOA and its variants are particularly promising as they offer the prospect of quantum advantage for a range of practically important problems~\cite{wolsey2014CombinatorialOptimization}.
Recent research has highlighted significant challenges for training VQAs~\cite{larocca_diagnosing_2022, Ragone2024Lie}, including practically interesting instances of QAOA (\cite{allcock2024QAOADLA,kazi2024QAOADLA}). 
A primary challenge is optimizing the parameters of the PQC. In most proposed VQAs, if the parameters are randomly initialized it is exponentially likely that the initial parameters will fall in a part of the optimization landscape with exponentially vanishing variance, causing parameter optimization to fail~\cite{Larocca2024ReviewBP}. 

Several methods have been proposed to address these challenges, which aim to initialize the parameters of a variational quantum algorithm (VQA) potentially being close to an optimal set of parameters, thereby mitigating the barren plateau problem. In this context, Goh et al.~\cite{goh_lie-algebraic_2025} introduced a pretraining scheme for VQAs based on a Lie-algebraic classical simulation framework termed `$\mathfrak{g}$-sim'. Their method applies $\mathfrak{g}$-sim to efficiently obtain optimal parameters for a related, simulable problem, which are then used to initialize the full VQA for the target problem. Demonstrating this approach on QAOA for the MaxCut problem, Goh et al.~\cite{goh_lie-algebraic_2025} showed that Lie-algebraic pretraining can significantly outperform random initialization.

In parallel, Bennett et al.~\cite{bennett_non-variational_2024,bennett_analysis_2024} focused on enhancing the quantum walk-based optimization algorithm (QWOA), originally introduced by Marsh and Wang~\cite{marsh2019, marsh2020, marsh2021} as a generalization of the Quantum Approximate Optimization Algorithm (QAOA). QWOA leverages continuous-time quantum walks (CTQW)~\cite{QWbook2014} to explore the solution space more naturally and handle constraints with greater flexibility as demonstrated in~\cite{ bennett2021quantum, slate2021quantum, matwiejew2023, Qu2024}. 
Building on this foundation, Bennett et al. ~\cite{bennett_non-variational_2024,bennett_analysis_2024} proposed a non-variational variant of QWOA, referred to in this paper as NV-QWOA. Unlike conventional variational algorithms that involve optimizing a large number of parameters, NV-QWOA reduces the trainable parameters to just three hyperparameters governing the evolution time and walk dynamics.  
This streamlined approach eliminates the need for costly classical optimization loops, thereby improving scalability and robustness against barren~\cite{bennett_non-variational_2024, bennett_analysis_2024, bennett_maxcut2025}.

In this paper, we compare the heuristics and performance of Lie algebraic pretrained QWOA and NV-QWOA for the Maxcut problem. We compare the parameter space explored by each model and benchmark their performances on 200 3-regular graphs and 200 Erd\H{o}s–R\'{e}nyi (ER) graphs with 16 vertices. We find that Lie algebraic pretrained QWOA's unstructured parameter initialization is significantly outperformed by NV-QWOA. In particular, NV-QWOA achieves a mean approximation ratio of 98.9\% within 60 tuning iterations, while Lie algebraic pretrained QWOA reaches only 88.8\% at the maximum of 500 tuning iterations. That optimization proceeds more quickly for NV-QWOA is not surprising given its significantly smaller parameter space, however, that an algorithm with so few tunable parameters reliably attains a near perfect approximation ratio after relatively few iterations is remarkable. 

\section{Quantum Algorithms for Combinatorial optimization problems}\label{sec:comb}
Combinatorial optimization problems consist of a discrete set of candidate solutions, which can be represented as a vector of \( s \) variables \( \mathbf{x} = (x_1, \ldots, x_s) \). The problem is specified by \( m \) clauses, each of which is defined on a subset of the variables. These clauses have an associated quality \( q_\alpha(\mathbf{x}) \) that is 1 if the solution \( \bm{x} \) satisfies the clause and 0 otherwise. The linear combination of these clauses defines a quality function, as given by
\begin{equation}
    q(\bm{x}) = \sum_{\alpha=1}^{m} w_\alpha q_\alpha(\bm{x}), \quad w_\alpha \in \mathbb{R},
    \label{eq:comb_cost}
\end{equation}
where a higher weight, $w_\alpha$, corresponds to a more important clause. Combinatorial optimization algorithms aim to find the solution string $\bm{x}$, for which the associated value is the maximum possible quality, $q_{\text{max}}$. The success of an algorithm is typically measured in terms of the approximation ratio, 
\begin{equation}
    A_r = \frac{q(\bm{x}^*)}{q_{\text{max}}}.
    \label{eq: approxratio}
\end{equation}
This value compares the quality of the identified solution $\bm{x}$* with the maximum quality. This is used as most problems of interest are impossible to exactly solve, so algorithms aim to approximate the solution and find a solution $\bm{x}$* which gives an approximation ratio close to one~\cite{wolsey2014CombinatorialOptimization}.

The main challenge in solving these problems is that the search space typically grows exponentially or super-exponentially with the problem size. This exponential growth will be illustrated in the following section, which introduces Maxcut, the combinatorial optimization problem which we will use to benchmark.

\subsection{Maxcut Problem Formulation}\label{sec:Maxcut}

\begin{figure}[h]
  \centering
  \bigskip
    \begin{subfigure}[b]{0.23\textwidth}
      \centering
      \begin{tikzpicture}[scale=0.6, every node/.style={circle, draw, minimum size=0.6cm}]
        \node (v1) at (0,0) {1};
        \node (v2) at (0,2) {2};
        \node (v3) at (0,4) {3};
        \node (v4) at (0,6) {4};
        
        \draw (v1) -- (v2) -- (v3) -- (v4);
      \end{tikzpicture}
      \medskip
      \caption{Path graph (\(P_4\))}
      \label{fig:path_graph}
    \end{subfigure}
  \hfill
  \begin{subfigure}[b]{0.23\textwidth}
    \centering
    \begin{tikzpicture}[scale=1, every node/.style={circle, draw, minimum size=0.6cm}]
      \foreach \i in {1,...,5}{
        \node (v\i) at (90 + 72*\i:1.5cm) {\i};
      }
      
      \foreach \i in {1,...,5}{
        \pgfmathtruncatemacro{\next}{mod(\i,5) + 1};
        \draw (v\i) -- (v\next);
      }
    \end{tikzpicture}
    \medskip
    \caption{Cycle graph (\(C_5\))}
    \label{fig:cycle_graph}
  \end{subfigure}
  \hfill
  \begin{subfigure}[b]{0.23\textwidth}
    \centering
    \begin{tikzpicture}[scale=1, every node/.style={circle, draw, minimum size=0.6cm}]
      \node (v1) at (0,0) {1};
      \node (v2) at (2,0) {2};
      \node (v3) at (1,1.732) {3}; 
      \node (v4) at (1, 0.577) {4}; 
      
      \draw (v1) -- (v2) -- (v3) -- (v1);
      \draw (v1) -- (v4);
      \draw (v2) -- (v4);
      \draw (v3) -- (v4);
    \end{tikzpicture}
    \medskip
    \caption{3-regular graph (\(G_4\))}
    \label{fig:3_regular_graph}
  \end{subfigure}
  \hfill
  \begin{subfigure}[b]{0.23\textwidth}
    \centering
    \begin{tikzpicture}[scale=1, every node/.style={circle, draw, minimum size=0.6cm}]
      \foreach \i in {1,...,5}{
        \node (v\i) at (90 + 72*\i:1.5cm) {\i};
      }
      
      \foreach \i in {1,...,5}{
        \foreach \j in {\i,...,5}{
          \ifnum\i<\j
            \draw (v\i) -- (v\j);
          \fi
        }
      }
    \end{tikzpicture}
    \medskip
    \caption{Complete graph (\(K_5\))}
    \label{fig:complete_graph}
  \end{subfigure}
  
  \caption{Examples of common graph structures.}
  \label{fig:graph_structures}
\end{figure}
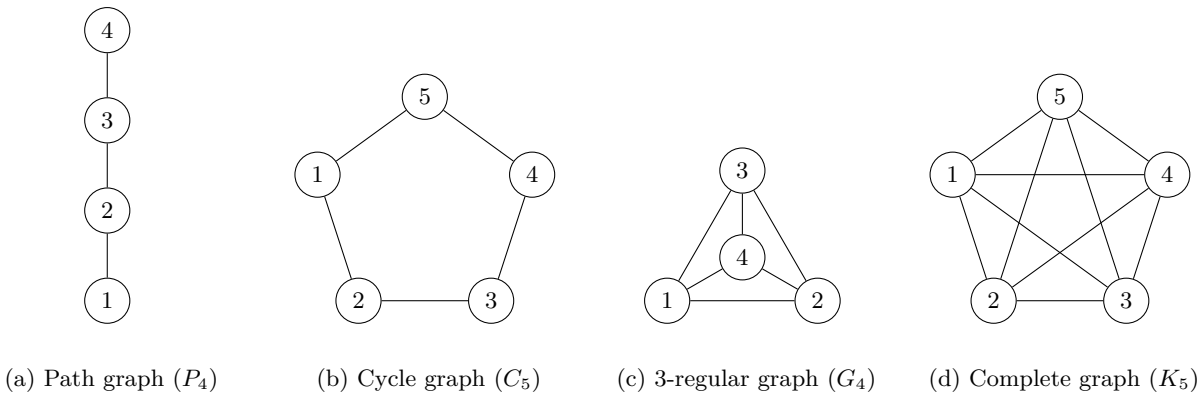

Maxcut is a longstanding benchmark for combinatorial optimization algorithms and  continues to serve as a primary benchmark for QAOA and its variants~\cite{farhi2014qaoa,blekos2024reviewQAOA,bennett_non-variational_2024}. 
To formulate Maxcut, we first recall that a graph $G$ is an ordered pair $(V,E)$,
where $V$ is the set of vertices and $E$ is the set of edges, with
$E \subseteq \{ \{u,v\} \mid u,v \in V,\ u \neq v \}$. For example, the path graph with 4 vertices may be expressed as \( V = \{v_1, v_2, v_3, v_4\} \) and \( E = \{(v_1, v_2), (v_2, v_3),(v_3, v_4)\} \), indicating that only sequential vertices are connected. This graph and other common graphs are seen in \Cref{fig:graph_structures}.

The goal of Maxcut is to partition a target graph's vertices into two subsets which has the maximum number of edges with end vertices in the opposite subset. Such a partition is called the graph's `Maxcut'. Thus, the solution space is the set of all partitions which labels vertices $v_i$ into two distinct subsets. These solutions can be represented as bit strings $\bm{x}=\{x_i\}$ with $x_i \in \{0,1\}$. Thus, the computational basis of $n$ qubits represents the solution space to the Maxcut problem for a graph with $n$ vertices, noting that each cut is double counter through global bit-flip symmetry (e.g. $00101 \equiv 11010$). The quality of a cut $\bm{x}$ is given by
\begin{equation}
    q(\bm x) = \sum_{\{i,j\} \in E} ( x_i-x_j )^2,
    \label{eq: Maxcutcost}
\end{equation}
since $\bm{x}$ cuts the edge $\{i,j\}$ if and only if $i$ and $j$ are in different sets under the cut.

The exponential nature of this problem is clear, with $2^{n-1}$ candidate solutions for a graph with $n$ vertices. This exponential growth makes solving Maxcut on large unstructured graphs impossible to solve exactly, so approximate optimization algorithms are required. The best known classical algorithm that runs in polynomial time and can be applied to any graph is the Goemans–Williamson (GW) algorithm, with a guaranteed approximation ratio of 87.86\%~\cite{goemans1995GW}. For 3-regular graphs, where every vertex is incident with exactly three edges (for example, Figure~\ref{fig:3_regular_graph}), a higher minimum approximation ratio of 93.26\% is guaranteed~\cite{halperin20043regClassical}. Although these algorithms can find very good solutions; in applications of Maxcut, such as circuit designs, even the smallest improvements can have large impacts~\cite{commander2009maximumCut}.

\subsection{Quantum Walk Optimization Algorithm (QWOA) for Maxcut}\label{sec:QWOA}
QAOA is a foundational quantum algorithm for combinatorial optimization problems which represents candidate solutions of a $n$ vertex graph as a $n$-bit string $\bm{x}$ using the computational basis states. The standard Maxcut quality function \eqref{eq: Maxcutcost} is than mapped similarly to an observable $Q$ defined as
\begin{equation}
    Q = \sum_{\{i,j\} \in E} \frac{1}{2} (\mathbb{I} - Z_i Z_j),
    \label{eq: Maxcut Hc}
\end{equation}
where $Z_i$ denotes the Pauli Z operator acting on the $i$-th qubit. This preserves the quality function \eqref{eq: Maxcutcost}, with the expectation value of a candidate solution with respect to Q, $\bra{\bm{x}}Q\ket{\bm{x}}$, being equivalent.

To find a high quality solution to to the target problem QAOA is applied to an initial state given by 
\begin{equation}
\ket{s}=\frac{1}{\sqrt{N}}\sum_{\bm x\in S} \ket{\bm x}
\end{equation}
which has probability equally spread between candidate solutions. This initial state is than acted upon by alternating between adding a phase to each basis state $\ket{\bm x}$ proportional to $q(\bm x)$, and performing a continuous-time quantum walk (CTQW) on an adjacency matrix $M$ for $p$ iterations. The CTQW acts to spread amplitude between solutions linked by the specified adjacency matrix, referred to as the mixer. The mixer acts to spread the probabilities between similar solutions allowing for constructive interference to increase the probability of measuring more favorable solutions. These steps give rise to the PQC 
\begin{equation}\label{eq:QWOA-PQC}
U(\bm\gamma,\bm t) = \prod_{k=1}^p e^{-iM\gamma_k} e^{-iQt_k} ~
\end{equation}
where $p\in\mathbb{N}$ is the depth, where $\bm \gamma,\bm t\in \mathbb{R}^p$ are lists of tunable parameters called phase separation coefficients and mixing times respectively, $Q$ is given by \eqref{eq: Maxcut Hc}, and $M$ is the mixer. The Quantum Walk-based Optimization Algorithm (QWOA) generalizes QAOA by defining the mixed based on the target problem to more effectively spread the probability to higher quality solutions. The reader is referred to \cite{marsh2020} for a discussion of different mixers. For this paper we only consider the problem of Maxcut using the `binary' or `hypercube' mixer defined as
\begin{equation}\label{eq:bin-mixer}
M = \sum_{j=1}^n  X_j ~,
\end{equation}
where $X_i$ denotes the Pauli-X operator acting on the $i$th qubit. For this paper QWOA will always refer to QWOA using the binary mixer, which is identical to standard QAOA.\\\\QWOA works by randomly initializing $(\bm{\gamma}, \bm{t})$ and then iteratively tuning them using a classical optimization routine which takes the quality of the evolved state, given by
\begin{equation}
    q(\bm{\gamma}, \bm{t}) = -\text{Tr}\left( Q U(\bm{\gamma}, \bm{t})\ket{s}\bra{s} U^\dagger(\boldsymbol{\gamma}, \bm{t}) \right),
    \label{eq: QAOA costfunction}.
\end{equation}
as input and outputs an improved set of parameters $(\bm{\gamma^*}, \bm{t}^*)$. Successful classical parameter optimization terminates upon optimal parameters $(\bm \gamma^*,\bm t^*)$ such that there is a high probability of measuring a near-optimal solution to the combinatorial optimization problem. 

The performance of QWOA is known to improve with increasing PQC depth $p$, and in some cases a minimum depth of $p>\mathcal{O}(\log(n))$ is required to allow the possibility of outperforming classical algorithms~\cite{farhi2014qaoa,hastings2019qaoa,bravyi2020obstaclesQAOA,farhi2020qaoaTypical,akshay2022QAOAdepthreq}. Unfortunately, deep practically important instances of QWOA suffers from a concentration of exponentially many parameter assignments to a single, sub-optimal quality, resulting in a region of near-zero gradient dominating the optimization landscape \cite{allcock2024QAOADLA,kazi2024QAOADLA,Ragone2024Lie}. This phenomenon is referred to as a barren plateau and means that optimization algorithms are rarely able to proceed from random initial parameters to optimal parameters, but instead get stuck on this plateau. This exponential concentration effects all types of optimization algorithm and poses a fundamental barrier to implementation \cite{Larocca2024ReviewBP}. We will now consider two contrasting recently proposed methods for improving the performance of QWOA: Lie algebraic pretraining and the non-variational QWOA.

\section{Lie algebraic pretraining For Maxcut}\label{sec:gsim}
Originally proposed by Huggins \textit{et al}.\ \cite{huggins2019firstSynergy} in 2019, classical pretraining strategies for VQAs have recently been gathering increased attention~\cite{Larocca2024ReviewBP}. Classical pretraining works by first classically solving a simplified auxiliary problem which approximates the target problem. It must be possible to then directly identify a set of parameters that initialize a PQC on the quantum device. Then additional gates are added, giving the required expressivity required to solve the complete target problem using the standard VQA framework. These additional gates are initialized to the identity (by setting their parameters to zero) to preserve the classically solved solution during the first run on the quantum device. This ensures that the full algorithm starts with the classically solved solution, even with the additional expressive power. This framework has been used with a variety of classical simulation methods depending on the VQA targeted for pretraining (for example ~\cite{goh_lie-algebraic_2025,rudolph2023tensorpretraing,dborin2022matrixPretrain}). Here we consider the Lie algebraic pretraining method applied to QWOA as proposed by Goh \textit{et al.}~\cite{goh_lie-algebraic_2025}, which makes use of Lie algebraic classical simulation. In this section the fundamentals of Lie algebras for VQAs is presented before introducing Lie algebraic pretraining for QWOA.

\subsection{Dynamical Lie algebras (DLAs)}\label{sec:DLA}
The dynamical Lie algebra (DLA), henceforth denoted $\mathfrak{g}$, is the Lie algebra associated with a quantum circuit, and allows us to quantify how expressive a general circuit is. To formally define it we first define the general form of a PQC as 
\begin{equation}
    U(\theta) = \prod_{p=1}^{P} \prod_{k=1}^{K} e^{-i \theta_{pk} H_k}.
\end{equation} 
Given such a circuit, the associated DLA is defined as the real vector space given by the span of all possible nested commutators of \(\{iH_1, \dots, iH_K\}\). This is defined as
\begin{equation}
    \mathfrak{g} = \text{span}_{\mathbb{R}} \left\langle \{iH_1, \dots, iH_K\} \right\rangle_{\text{Lie}} \subseteq \mathfrak{su}(2^n),
    \label{eq:DLA}
\end{equation}
where \(\left\langle \cdots \right\rangle_{\text{Lie}}\) denotes the set generated by taking repeated commutators until no new linearly independent element is found.

Knowledge of the DLA of a PQC allows one to infer which unitaries it is capable of expressing. With $\mathfrak{g}$ as above, let $G = e^\mathfrak{g}\defeq \{ e^H : H\in\mathfrak{g} \}$, the Lie group generated by $\mathfrak{g}$.
The elements of $G$ are precisely the unitaries that can be implemented by a series of gates with generators in $\mathfrak{g}$ and some assignment of parameters; i.e.\ unitaries of the form
\begin{equation}
    U(\theta) = \prod_{p=1}^{P} \prod_{k=1}^{K} e^{-i \theta_{pk} H_k} = e^{g(\theta)}.
    \label{eq:PQCfromDLA}
\end{equation}
This holds for arbitrary circuit depths $P \in \mathbb{N}$ and parameter sets: $\theta = \{\theta_{pk}\} \in \mathbb{R}^{P\times K}$. If the PQC can express the full space of unitaries U($2^n$) then its DLA is all of $\mathfrak{su}(2^n)$ and has a dimension exponential in the number of qubits. Hence, the dimension $\mathfrak{g}$ as a real vector space may be related to the expressivity of the PQC to directly show that if the dimension of the PQC's DLA grows exponentially with $n$ the circuit will suffer from the barren plateau~\cite{fontana2024Adjointbarren,Ragone2024Lie}. 

VQAs which are free from barren plateaus also have DLAs which scale polynomially with the number of qubits. Therefore, by mapping the circuit into this polynomial space, circuits whose DLA-dimension scales only polynomially in the system size $N$ can be efficiently classically simulated~\cite{somma2005gsim1,somma2006gsim2}. This simulation framework is detailed in Appendix~\ref{ap:Lie algebra sim math}. 

Recently, Goh \textit{et al.}\ \cite{goh_lie-algebraic_2025} have reformulated this classical simulation strategy into a framework for simulating, optimizing, and training quantum circuits. They have also proposed a new pretraining scheme for VQAs, which we call `Lie algebraic pretraining'.

\subsection{Lie algebraic pretraining}\label{sec:gsim pretraing}
Here we introduce the new Lie algebraic pretraining method due to Goh \textit{et al.}\ \cite{goh_lie-algebraic_2025}. Consider a VQA for which there exists a problem instance whose associated PQC has an exponentially growing DLA and so suffers from a barren plateau, and a secondary problem instance with a polynomial DLA-dimension and which is efficiently trainable and efficiently classically simulable using $\mathfrak{g}$-sim. With the large-DLA problem instance as the target problem and the small-DLA problem instance as an auxiliary problem, we apply Lie algebraic pretraining as follows:
\begin{enumerate}
    \item Use Lie algebraic simulation to efficiently classically simulate the PQC of the auxiliary problem, employing a circuit depth sufficient for over-parameterization (i.e.\ equal to $\dim(\mathfrak{g})$;\cite{Larocca2023Overparametrization}, we optimize the parameters of the auxiliary PQC and converge to the global optimum \cite{goh_lie-algebraic_2025}.
    \item Augment the PQC of the auxiliary problem with gates from the target problem's PQC.
    \item Run a VQA using this new PQC, initializing the target problem's gate parameters to $0$ and initializing the auxiliary problem's gate parameters to the optimal parameters that were found in Step 1 using Lie algebraic classical simulation.
\end{enumerate}
Steps 2 and 3 aim to transfer the pretrained parameters into the optimization landscape of the target problem.

In line with the general strategy outlined above, Lie algebraic pretraining for QWOA for Maxcut is performed by first classically optimizing QWOA on an auxiliary graph which is associated with a small-DLA QWOA PQC. This produces a set of parameters, $\bm{\gamma^A},\bm{t}$, so that measurement of $\ket{\bm{\gamma^A},\bm{t}}$ has a high probability of yielding the state encoding a maximum-size cut of the auxiliary graph.

Upon determining parameters $\bm{\gamma^A},\bm{t}$ which give the Maxcut on the auxiliary graph, additional expressivity is added to the PQC to allow for the target graph to be solved, yielding the modified PQC given by
\begin{equation}\label{eq:gsimQWOA-PQC}
    U(\bm{\gamma^A}, \bm{\gamma^T},\bm{t}) = \prod_{k=1}^p e^{-i t_k M} e^{-i \gamma_k^A Q_A} e^{-i \gamma_k^T Q_T},
\end{equation}
where $\bm{\gamma^A},\bm{t}$ are the optimal parameters found for the auxiliary graph, $\bm{\gamma^T}$ are additional parameters associated with the quality operator $Q_T$ of the target graph, $Q_A$ is the quality operator of the auxiliary graph, and $M$ is the binary mixer (see \eqref{eq:bin-mixer}).

The additional parameters $\bm{\gamma^T}$ are all initialized at zero so that the new gates generated by $H_T$ are identities, ensuring that the first iteration of the VQA using the above modified PQC is the same as the standard QWOA for the auxiliary graph with its optimal parameters $\bm{\gamma^A},\bm{t}$. All parameters $\bm{\gamma^A},\bm{\gamma^T},\bm{t}$ are then varied to optimize the expectation value of $Q_T$, i.e.\ the quality function of QWOA for Maxcut on the target graph.

There are very few graph structures that are known to yield a QWOA for Maxcut with a polynomial DLA. Indeed, the known structures are all very simple and have no clear relation to graph structures of interest~\cite{allcock2024QAOADLA}. Despite this, initial trials of Lie algebraic pretraining were conducted by Goh \emph{et al}.\ \cite{goh_lie-algebraic_2025}. These trials used the path graph as the auxiliary problem and 3-regular and ER random graphs with an edge probability of 30\% as the targets. The ER model generates random graphs by connecting each pair of vertices with a fixed probability. Simulations of both randomly initialized and Lie algebraic pretrained QWOA with a depth of $p=256$ were completed for both target graphs at 16 vertices. At this depth, and with non-trivial graph structures, QWOA is known to exhibit a barren plateau~\cite{larocca_diagnosing_2022,kossmann2022deepQAOA}. Across 200 trials of each graph type, QWOA using Lie algebraic pretraining outperformed random initialization for a majority of instances. Furthermore, QWOA with pretraining was able to achieve approximation ratios surpassing the GW threshold in 72.5\% of 3-regular graphs and 66.5\% of ER graphs, after 500 iterations of the L-BFGS optimization algorithm. This contrasts with random initialization which, suffering from a barren plateau, achieved approximation ratios substantially lower than the GW threshold for all trials. This was particularly notable as the path graph, Figure \ref{fig:path_graph}, has no clear relation to 3-regular, Figure \ref{fig:3_regular_graph}, or ER graphs, both of which are substantially more complicated. 

\section{Non-variational QWOA}\label{sec:NV-QWOA}
NV-QWOA was proposed by Bennett \textit{et al.}\ \cite{bennett_non-variational_2024,bennett_analysis_2024} and has already seen remarkable success in a range of combinatorial optimization problems \cite{bennett_non-variational_2024}. It is characterized by parameterizing the $2p$ parameters of the QWOA PQC in terms of just $3$ hyperparameters: $\beta\in (0,1)$, $\gamma,t\in\mathbb{R}$. Namely, the phase separation coefficients and mixing times are chosen to respectively increase and decrease over the domain $\left[\beta\gamma/\sigma, \gamma/\sigma\right]$ (resp.\ $[\beta t,t]$), usually in a linear manner. Here $\sigma$ denotes the standard deviation of the objective function of the combinatorial optimization problem of interest, a quantity that may be estimated from random sampling. Hence, the PQC for NV-QWOA is simply 
\begin{equation}\label{eq:NV-QWOA-QPC}
U(\beta,\gamma,t) = \prod_{k=1}^p e^{-iM\gamma_k} e^{-iQt_k} ~,
\end{equation}
where
\begin{align}
\gamma_k &= \frac{\gamma}{\sigma}\left(\beta + \frac{k-1}{p-1}(1-\beta)\right) ~, \\ 
t_k &= t\left(1 + \frac{k-1}{p-1}(\beta-1) \right) ~,
\end{align}
and $\sigma$ is the standard deviation (which may be approximated via random sampling) of the quality function of the combinatorial optimization problem of interest. The three hyperparameters are then optimized in the same way as the gate parameters are optimized in a standard VQA.

The radical restriction of QWOA's $2p$-dimensional parameter optimization solution space to a $3$-dimensional space in NV-QWOA allows even randomly initialized parameter optimization to converge quickly.

It should be noted that NV-QWOA imposes additional restrictions on the mixer to ensure efficient performance, and these form one of the major contributions of Bennett \textit{et al.}. We state them here without explanation and refer the interested reader to Section II.C of Bennett \textit{et al.} \cite{bennett_non-variational_2024}. Consider a combinatorial optimization problem with solution space $S$ and quality function $q$. First, for every solution $\bm{y}\in S$ and for small $t\in \mathbb{R}_{>0}$ we require that
\begin{equation}\label{eq:nec.1}
   e^{-i Mt} \ket{\bm{y}} = \sum_{\delta=0}^D \left( e^{-i \delta \phi(t)} \sum_{\bm{x}\in S} r_{\bm{x}} (t) \ket{\bm{x}}\right), 
\end{equation}
where $D$ is the diameter of the mixing graph, $\phi$ is a function $\mathbb{R}_{>0} \longrightarrow (0,\pi)$, and $r_{\bm{x}}$ is a function $\mathbb{R}_{>0}\longrightarrow \mathbb{R}_{>0}$ for every $\bm x\in S$, and for all $\delta\in\{0\}\cup\mathbb{N}$ we define
\begin{equation}
    \delta_{\bm{y}} \defeq \{ \bm{x} : d(\bm{x},\bm{y}) = \delta \}~,
\end{equation}
where $d(\bm{x},\bm{y})$ is the distance between $\bm{x}$ and $\bm{y}$ on the mixing graph. In other words, the mixing unitary $e^{-iMt}$ acts on $\ket{\bm{y}}$ to produce a superposition of all states in $S$, each with a phase proportional to its distance from $\bm{y}$ in the mixing graph.
Second, for every solution state $\bm{x}$ we require that
\begin{equation}\label{eq:nec.2}
    (\mu_{\delta_{\bm{x}}} - q(\bm{x})) \approx -\alpha_{\delta} (q(\bm{x}) - \mu)
\end{equation}
where $\mu_{\delta_{\bm{x}}}$ is the mean of $q$ over $\delta_{\bm{x}}$, $\mu$ is the mean of $q$ over the solution space $S$ (i.e.\ the global mean of $q$), and $\alpha_\delta \in \mathbb{R}$ increases monotonically with $\delta$ (at least up to some considerable $\delta$). This says that the difference of $q(\bm{x})$ and the mean of $f$ over $\delta_{\bm{x}}$ is roughly proportional to the difference of $q(\bm{x})$ and the global mean of $q$. Bennet \textit{et al.}\ \cite{bennett_non-variational_2024} have shown that these conditions are satisfied by many commonly encountered mixers, including the binary mixer (in \cref{eq:bin-mixer}) used for Maxcut.

As recently highlighted by Cerezo \textit{et al.}\ \cite{cerezo_does_2025}, all known randomly initialized VQAs that are free from barren plateaus also can be efficiently classically simulated. Due to its small parameter space, NV-QWOA appears to side-step the barren plateau phenomenon, while maintaining sufficient expressivity to reach the solution.

\section{Methodology}
To compare the performance of standard QWOA with Lie algebraic  pretrained QWOA and NV-QWOA, all three algorithms were applied to Maxcut on the same sets of graphs. Following the methodology of Goh \textit{et al.} \cite{goh_lie-algebraic_2025}, 400 graphs were generated for the purpose of these simulations. Namely, 200 3-regular graphs and 200 ER random graphs with edge probability $0.3$ (ER(0.3)) were generated, each on 16 vertices. The depth of the PQC was set to 256 for all three of the algorithms (i.e.\ $p= 256$ in \cref{eq:QWOA-PQC,eq:NV-QWOA-QPC,eq:gsimQWOA-PQC}). This relatively large depth is a requirement of the Lie algebraic  pretrained QWOA as it ensures convergence on a global minimum through reparameterization~\cite{goh_lie-algebraic_2025}. The metric used for the comparison between the algorithms was the approximation ratio of the final expectation value of the algorithm against the maximum possible cut. The parameters were restricted to the interval $[0,2\pi]$, as is typical~\cite{blekos2024reviewQAOA}. The NV-QWOA was run with all 3 hyperparameters randomly initialized for each graph. Code was developed to simulate the quantum circuit using the Python package Qiskit to simulate the quantum circuits and the L-BFGS through Scipy’s minimize function for optimization~\cite{qiskit2024,2020SciPy-NMeth}. To ensure efficiency, the gradients were calculated according to the procedure presented by \textit{Jones et al.}\cite{jones2020efficientGrad}. All simulations were performed on the Setonix supercomputer, hosted by the Pawsey Supercomputing Centre.

Since the path graph was used as the auxiliary problem for pretraining and is independent of the target problem, pretraining needed to be performed only once. Pretraining QWOA for the path graph was done by randomly initializing parameters and optimizing until convergence to the exact solution. This pretraining was conducted solely to simulate the full QWOA on a classical device, so an efficient implementation with Lie algebraic simulation was unnecessary. Hence, the pretraining was completed with the same Qiskit code used for the full QWOA simulation runs.

\section{Results and discussion}\label{sec:results}
\subsection{Initial Parameters from Pretraining}
\begin{figure}[h!]
    \centering
    \includegraphics[width=\textwidth]{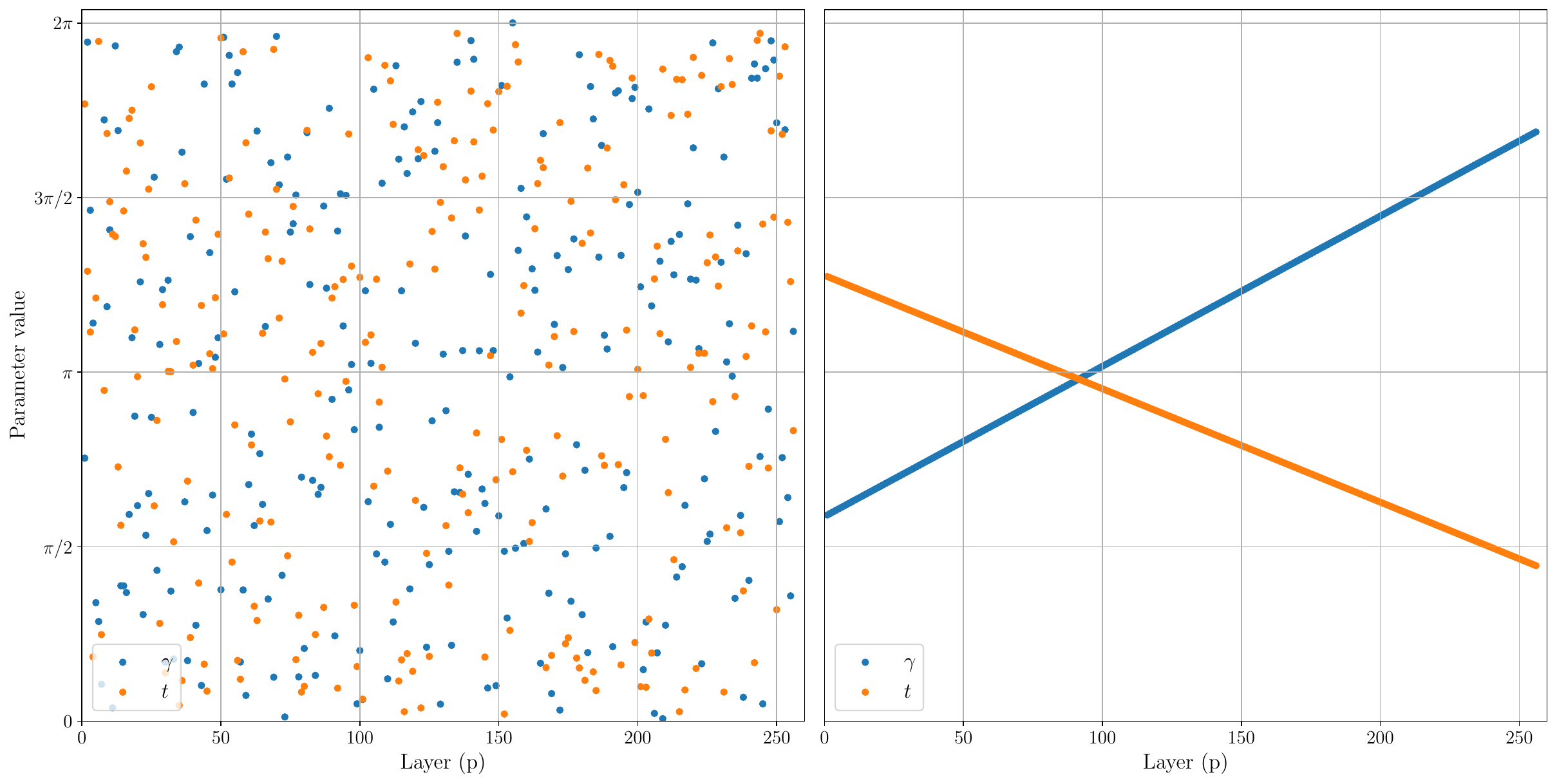}
    \caption{Initial parameters for QWOA for Maxcut with 256 layers ($p=256$). (A) Example solution parameters for the 16-vertex path graph, used as initial parameters for Lie algebraic pretraining. (B) Example parameters for NV-QWOA, namely $(\beta,\gamma,t) = (0.35,5.3,4)$.}
    \label{fig:initial_params}
\end{figure}

The Lie algebraic pretraining on the path graph with 16 vertices was successfully completed, converging to an approximation ratio of 99.9999996\%, with 99.999998\% of the amplitude being in the optimal partition. The initial parameters determined by the Lie algebraic pretraining are unstructured and clearly unrelated to the linear grading used by the NV-QWOA as seen in Figure \ref{fig:initial_params}. The distinct differences in parameter sets emphasize that the Lie algebraic pretraining method is not inadvertently improving training through a similar mechanism as the NV-QWOA.

\begin{figure}[h!]
    \centering
    \includegraphics[width=\textwidth]{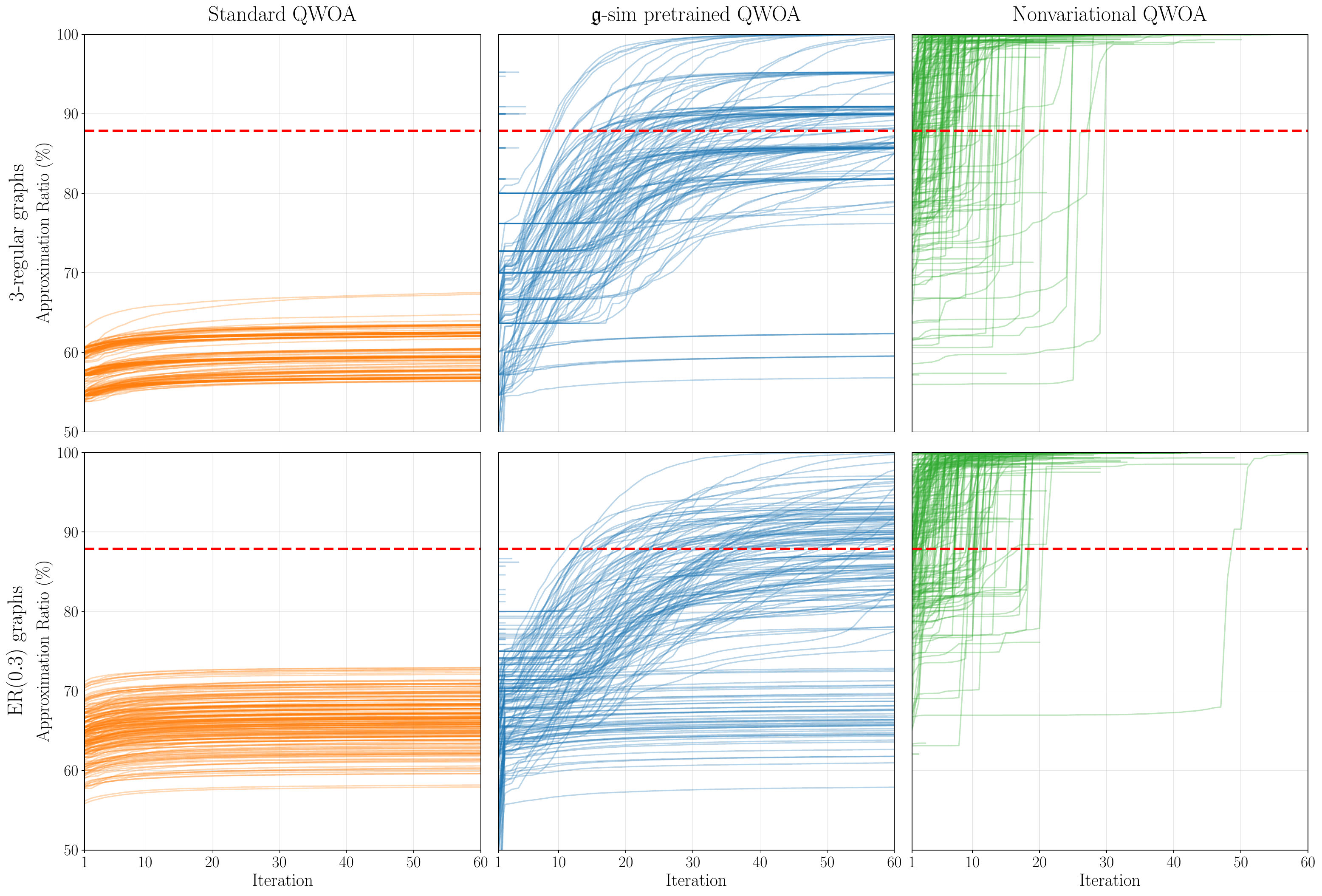}
    \caption{Individuals runs of the same problem sets on three different algorithms.}
    \label{fig:individual_runs}
\end{figure}

\begin{figure}[h!]
    \centering
    \includegraphics[width=0.67\textwidth]{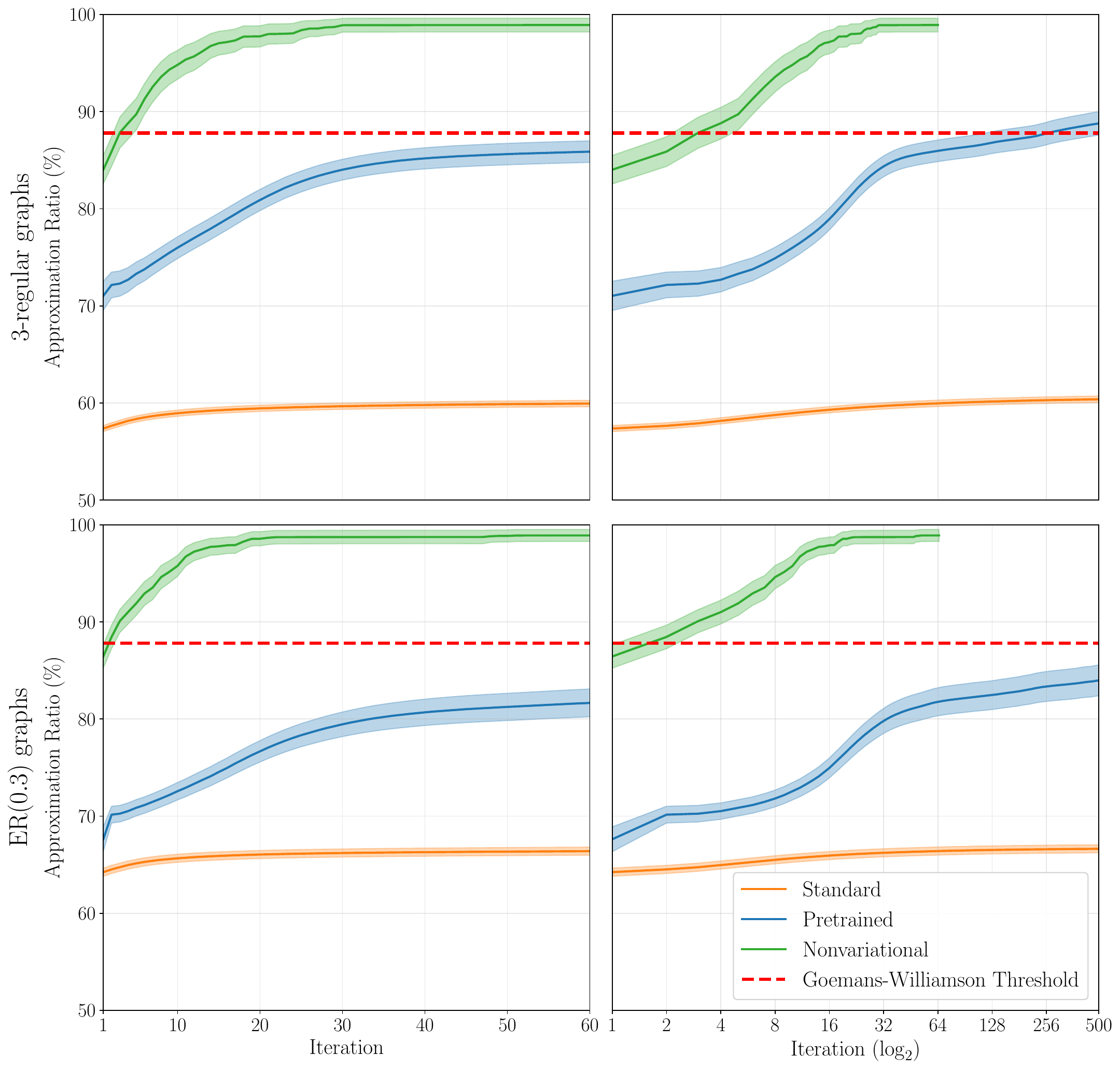}
    \caption{Mean approximation ratio (with 95\% confidence interval) of each iteration across runs of the same problem sets on the three algorithms.}
    \label{fig:mean-runs}
\end{figure}
The performances of QWOA with random parameter initialization and Lie algebraic pretraining and of NV-QWOA are shown in Figure \ref{fig:individual_runs}. For the 200 sampled 3-regular graphs, QWOA with Lie algebraic pretraining achieves a mean approximation ratio of 88.79\% after 500 iterations, while random initialization only reaches 60.38\% and only marginally improves upon random initialization. For the ER graphs, the mean approximation ratio is 83.97\% with Lie algebraic pretraining and 66.63\% with random initialization. The NV-QWOA achieves a mean approximation of greater than 98.9\% on both graph types. These mean performances are shown in Figure \ref{fig:mean-runs} where it is also seen that NV-QWOA consistently reaches convergence by at most 60 iterations.

Three broad trends emerge when considering the individual training history of each of the 400 trials using Lie algebraic pretraining. First, numerous instances of Lie algebraic pretraining fails to improve upon the performance level of the randomly initialized QWOA. Specifically, 3.5\% of the 3-regular graphs and 23.5\% of the ER graphs have final approximation ratios within 1\% of those achieved by their corresponding randomly initialized trials. A second minority of instances become trapped in a local minimum almost immediately, with fewer than 20 updates and improvement upon the initial approximation ratio of less than 5\%. This behavior is observed in 30\% of the 3-regular instances and 16\% of the ER instances. The NV-QWOA was substantially more consistent in its behavior, with every instance of NV-QWOA improving upon the standard QWOA. For NV-QWOA, 26 and 37 instances of 3-regular and ER graphs respectively got stuck in local minima. However, only 7 instances of each type that got stuck in local minima had final approximation ratios less than 95\%.

For Lie algebraic pretrained QWOA, instances that achieved a final approximation ratio surpassing the GW threshold comprised 67\% of 3-regular graphs and 49.5\% of ER graphs. Moreover, 7.5\% of 3-regular and 3\% of ER graphs converge on an approximation ratio greater than 99.99\%. The NV-QWOA surpassed the GW-threshold for 97.5\% and 98\% GW-threshold for 3-regular and ER graphs respectively. Remarkably 58.5\% of 3-regular and 60\%  of ER graphs converged on an approximation ratio greater than 99.99\%. Furthermore, this impressive convergence was usually achieved within just 20 iterations, while many instances of the Lie algebraic pretraining method failed to converge even after 500 iterations. This speaks to the power of limiting to just 3 hyperparameters to optimize over, as opposed to the more than 500 parameters used by both the standard and Lie algebraic pretraining strategy.

The consistent performance of NV-QWOA is impressive, particularly when considering that the 3 hyperparameters were randomly initialized for each run. Previous results showed that for a given parameter set the performance of QWOA with the hypercube mixer applied to Maxcut is relatively agnostic to specific graph instances. Specifically, a good parameter set for one graph instance will work equally well on another instance of the same graph type~\cite{brandao2018qaoafixed}. Indeed, there is a fixed set of suggested hyperparameters suggest by Bennett \textit{et al.}\ \cite{bennett_non-variational_2024}, however these were not used in this analysis. The ability for NV-QWOA to perform so consistently even without fixed hyperparameters suggests that its success is not dependent on an optimization landscape specific to Maxcut.

Notably, the fraction of instances that improved upon the GW threshold is 5.5\% and 17\% lower for 3-regular and ER graphs, respectively, compared to the results reported by Goh \emph{et al.}\ \cite{goh_lie-algebraic_2025}. Using the standard normal (z-score) approximation, a 95\% confidence interval was calculated for the proportion of instances surpassing the GW threshold in the population, with results indicating intervals between 60.48\% and 73.52\% for 3-regular graphs, and 42.57\% and 56.43\% for ER graphs. This points to the discrepancy going beyond sampling error for ER graphs, with the calculated 95\% confidence interval having an upper bound that is 10\% lower than the previously reported value. We hypothesis that this discrepancy is due to the pretraining step, which depends on selecting a random set of initial parameters and then optimizing them to find the optimal partition of the path graph. The combination of a deep QWOA circuit with the simplicity of the path graph results in multiple sets of parameters producing the same solution. These parameter sets all yield the Maxcut upon initializing the full QWOA for the path graph but may be in radically different locations to one another in the parameter landscapes of other graphs, hence leading to varying performance. This phenomenon may be investigated in future work to determine the extent of variation. This variation coupled with the performance gap points to NV-QWOA substantially outperforming Lie algebraic pretraining methods for QWOA on Maxcut. Nevertheless, Lie algebraic pretraining may still play a valuable role training VQAs for other applications, while NV-QWOA is limited to combinatorial optimization problems.

\section{Conclusions}\label{sec:conc}
In this work, we have compared the performance of randomly initialized QWOA, Lie algebraic pretrained QWOA, and NV-QWOA for Maxcut. Our results highlight the need for smart parameter initialization strategies. As expected, randomly initialized QWOA failed to find a good solution for every graph instance due to the barren plateau. Lie algebraic pretraining relies on the generalized heuristics of simply trying to initialize at a location in the optimization landscape away from the barren plateau. This succeeded on average and strongly improved upon random initialization. However, it failed to outperform the algorithm specific heuristics and structured parameter initialization of NV-QWOA.

In particular, NV-QWOA consistently converged to a mean approximation ratio of 98.9\% within $60$ iterations. While Lie algebraic pretrained QWOA achieved a mean approximation ratio of only 77.71\%, even after $500$ iterations. Furthermore, NV-QWOA failed to exceed the GW threshold of 87.9\% for less than 3\% of all graph instances. This consistency goes beyond previous results showing that a fixed parameter set leads to consistent performance agnostic to graph instance as the 3 hyperparameters were randomly initialized. While this points to strong promise on other problems, further work is required to  evaluate the scalability of the proposed approach to significantly larger problem instances and to determine its generalizability across classes of combinatorial optimization problems. 

\section*{Acknowledgements}

This project was supported by the Australian Government via the Critical Technologies Challenge Program (CTCP) and the Advanced Strategic Capabilities Accelerator (ASCA). Substantial computational resource was provided by the Pawsey Supercomputing Research Centre. MZ wishes to acknowledge Josh Green and Matthew Goh for helpful discussions. JJ and JBW also acknowledge Mehul Pandita, Luke Antoncich, John Tanner, Edric Matwiejew, and Aidan Smith for their support with large-scale simulations, and Tavis Bennett and Mark Reynolds for valuable discussions. 

\section*{Data Availability}\label{sec:data availability}
All code and data produced for this work is available upon request.

\bibliography{bib}

\appendix

\section{Lie algebraic classical simulation}\label{ap:Lie algebra sim math}
To simulate quantum algorithms Lie algebraic simulation project from the full Hilbert space to a dim($\mathfrak{g}$) vector space~\cite{somma2005gsim1,somma2006gsim2}.
This is done using a map \(\Phi_{\text{ad}} : \mathfrak{g} \mapsto \mathbb{R}^{\dim(\mathfrak{g}) \times \dim(\mathfrak{g})}\) which takes elements of the DLA and returns a real valued matrix, called the adjoint representation.
For a DLA with a Schmidt orthonormal basis such that:
\[ \text{Tr}(B_iB_j)=\delta_{ij}~, \quad \forall B_i,B_j \in \mathfrak{g} ~, \] the adjoint representation is defined as Equation \ref{eq: adjoint rep H}. The values $f^\gamma_{\alpha \beta}$ are referred to as the structure constants of $\mathfrak{g}$, and are real valued as the trace of two skew Hermitian matrices yields real values.
\begin{equation}
    (\Phi^{\text{ad}}_\mathfrak{g}(iB_\gamma))_{\alpha \beta} \equiv f^\gamma_{\alpha \beta} = \text{Tr}\left( i B_\gamma \, [i B_\alpha, i B_\beta] \right) \quad f^\gamma_{\alpha \beta} \in \mathbb{R}
    \label{eq: adjoint rep H}
\end{equation}
From the linearity of the trace, and hence of the map $\Phi^{\text{ad}}_\mathfrak{g}$, the adjoint representation of any Hermitian operator which can be expressed as, $H = \sum_\alpha w_\alpha B_\alpha \in i\mathfrak{g}$, can be written in terms of the basis as
\begin{equation}
    \Phi^{\text{ad}}_\mathfrak{g}(H) = \sum_\alpha iw_\alpha\Phi^{\text{ad}}_\mathfrak{g}(iB_\alpha), \quad w_\alpha \in \mathbb{R}.
    \label{adjoint rep arb H}
\end{equation}
Here $i\mathfrak{g}$ denotes the DLA multiplied by the imaginary unit, corresponding to Hermitian operators. This map can also be applied to unitaries generated by the DLA, $\mathcal{G}= e^\mathfrak{g}$, as given by
\begin{equation}
    (\Phi^{\text{ad}}_\mathcal{G}(U=e^{iB_\gamma}))_{\alpha \beta} \equiv e^{\Phi^\text{ad}_\mathfrak{g}(iB_\gamma)} = e^{f^\gamma_{\alpha \beta}}.
    \label{eq: DLG representation}
\end{equation}
This representation describes exactly how an element of $\mathcal{G}$ evolves a basis element of $\mathfrak{g}$. This is given by
\begin{equation}
    U^\dag iB_\alpha U = \sum_\gamma^{\textrm{dim}(\mathfrak{g})} (\Phi^\text{ad}_\mathcal{G}(U))_{\alpha\gamma}iB_\gamma,\quad \forall U = e^\mathfrak{g},
    \label{eq: adjoint evolution}
\end{equation}
and generalises to any Hermitian operator O which is an element of $i\mathfrak{g}$.

Using the mathematical framework provided by the adjoint representation allows the cost function of a VQA to be rewritten in terms of the $\textrm{dim}(\mathfrak{g})$ vector space. By considering the evolution of the measurement observable O, and assuming O is supported by $\mathfrak{g}$ such that O $= \sum_\alpha w_\alpha B_\alpha \in i\mathfrak{g}$, Equation \ref{eq: adjoint evolution} may be used. This yields a cost function given by
\begin{equation}
    C_{\boldsymbol{\theta}}(O,\rho_{0}) = \sum_\alpha^{\textrm{dim}(\mathfrak{g})} \sum_\gamma^{\textrm{dim}(\mathfrak{g})} w_\alpha(\Phi^\text{ad}_\mathcal{G}(U))_{\alpha\gamma}\text{Tr}\left(B_\gamma \rho_{0}  \right)
    \label{eq: gsim costfunction}
\end{equation} 
This equation provides the exact form of the cost function in a dim($\mathfrak{g}$), allowing classical calculation which scales as $\mathcal{O}(LK\dim(\mathfrak{g})^3)$.

There are several requirements for Equation \ref{eq: gsim costfunction} to be efficiently implementable. The first requirements are that \( \dim(\mathfrak{g}) \in \mathcal{O}(\text{poly}(n)) \) and the measurement observable of interest must be a linear combination of the basis of \(\mathfrak{g}\). A more challenging requirement is that a Schmidt-orthonormal basis of \( \mathfrak{g} \), must be known, and that their associated structure constants must be efficiently calculable. The final requirement is that the expectation values of the basis elements over the initial state must be known. As most initial states are the ground state the final requirement is typically satisfied~\cite{somma2005gsim1}. Calculating the structure constants efficiently can be more challenging, however recent work has shown that it can be done for a large class of VQC algorithms with polynomial sized DLAs~\cite{anschuetz2023strucutueconstantseffeceint}.

For further details mathematical details an interested reader may refer to references \cite{goh_lie-algebraic_2025,somma2005gsim1,somma2006gsim2}. For an example of rudimentary code implementing this simulation procedure, readers may refer to reference \cite{kottmann_gsim_2024}. 

\end{document}